\documentclass[prl,twocolumn,showpacs,amsmath,superscriptaddress]{revtex4-1}
\usepackage{graphicx}% Include figure files
\usepackage{epsfig}
\usepackage{amsmath}
\usepackage{amssymb}
\usepackage{textcomp}
\usepackage{braket}
\usepackage{array}
\usepackage[squaren]{SIunits}  %Hk, musste ich ändern damit es kompiliert.
\usepackage{color}
\usepackage{dcolumn}% Align table columns on decimal point
\usepackage{hyperref}
\usepackage{breakurl} %for supplemental material link
\usepackage{tabularx}

\usepackage{graphicx}% Include figure files
\usepackage{dcolumn}% Align table columns on decimal point
\usepackage{bm}% bold math

\def\ket0{$\left|0\right>$}
\def\ket1{$\left|1\right>$}
\def\ca40{$^{40}\mathrm{Ca}^+$}
\def\T2{$\mathrm{T_2$}}
\def\Pr3{$\mathrm{Pr^{3+}}$}
\def\ket#1{$\left|#1\right>$}

\begin{document}

\newcommand*{\MAINZ}{QUANTUM, Institut f\"ur Physik, Universit\"at Mainz, Staudingerweg 7, 55128 Mainz, Germany}
\newcommand*{\ULM}{Institut f\"ur Theoretische Physik, and Center for Integrated Quantum Sciences and Technology, Universit\"at Ulm, Albert-Einstein-Allee 11, 89069 Ulm, Germany}
\newcommand*{\TELAVIV}{Raymond and Beverly Sackler School of Physics and Astronomy, Tel-Aviv University, Tel Aviv 69978, Israel}
\newcommand*{\JERUSALEM}{Racah Institute of Physics, The Hebrew University of Jerusalem, Jerusalem 91904, Givat Ram, Israel}

\title{Precise Experimental Investigation of Eigenmodes in a Planar Ion Crystal}

\author{H. Kaufmann}
\author{S. Ulm}
\author{G. Jacob}
\author{U. Poschinger}\affiliation{\MAINZ}
\author{H. Landa}\affiliation{\TELAVIV}
\author{A. Retzker}\affiliation{\JERUSALEM}
\author{M.B. Plenio}\affiliation{\ULM}
\author{F.~Schmidt-Kaler}\affiliation{\MAINZ}

\date{\today}% It is always \today, today, but any date may be explicitly specified

\begin{abstract}	
The accurate characterization of eigenmodes and eigenfrequencies of two-dimensional ion crystals provides the foundation for the use of such structures for quantum simulation purposes. We present a combined experimental and theoretical study of two-dimensional ion crystals. We demonstrate that standard pseudopotential theory accurately predicts the positions of the ions and the location of structural transitions between different crystal configurations. However, pseudopotential theory is insufficient to determine eigenfrequencies of the two-dimensional ion crystals accurately but shows significant deviations from the experimental data obtained from resolved sideband spectroscopy. Agreement at the level of 2.5$\times$10$^{-3}$  is found with the full time-dependent Coulomb theory using the Floquet-Lyapunov approach and the effect is understood from the dynamics of two-dimensional ion crystals in the Paul trap. The results represent initial steps towards an exploitation of these structures for quantum simulation schemes.
\end{abstract}

%options are
%   41.90.+e 	Other topics in electromagnetism; electron and ion optics (restricted to new topics in section 41)
%   45.50.Jf 	Dynamics and kinematics of a particle and a system of particles, Few- and many-body systems
%   37.10.Gh 	Atom traps and guides
%   37.10.Ty 	Ion trapping
%   03.67.-a    Quantum information,

\maketitle

% introduction
Accurate control of ion crystals is of major importance
for spectroscopy, quantum simulation, or quantum computing
with such experimental platform. Since the invention of
dynamical trapping by Paul \cite{paul1953neues}, this versatile instrument
has been adapted and optimized for specific purposes.
Charged particles, more specifically singly charged ions,
are confined in a radio frequency (rf) potential, which is
formed by tailored electrode structures. In the case of
the linear Paul trap, one aims for a quadrupole field
along one z axis, such that a harmonic pseudopotential in
x and y direction is formed. This radial potential strongly
confines the ions, while an additional weaker axial potential
in z direction is generated with static (dc) voltages
applied to end cap electrodes. Trapped ions are cooled
by laser radiation \cite{WinelandDehmelt1975} in the potential described by three
trap frequencies $\omega_{x,y,z}$ eventually forming a crystalized
structure.

The conditions of operation are characterized by two
anisotropy parameters where the radial confinement $\omega_{(x,y)}$
typically exceeds the axial dc confinement  $\omega_{z}$. For sufficiently
small values of $\alpha_{(x,y)} \equiv \omega^2_z/\omega^2_{(x,y)}$, the ion crystal is
linear and aligned along the weakest axis, the z trap axis;
all ions are placed in the node of the rf potential.
Spectacular highlights using linear crystals of cold ions
are the demonstration of quantum logic operations~\cite{cirac2000scalable,schmidt2003realization},
the generation of entangled states~\cite{roos2004control,MONZ}, sympathetic cooling
of ions of different species~\cite{barett2003,larson1986sympathetic}, or the quantum-logic
clock~\cite{PSCHMIDT}. To reach the level of quantum control, as
required in the experiments listed above, the first precondition
was a complete understanding of eigenmodes
and eigenfrequencies for such stored linear ion crystals~\cite{james1998quantum,steane,marquet2003phonon}.

For larger numbers of ions, or for larger values of $\alpha$, the
linear crystal undergoes a transition to a zigzag structure
and eventually to a fully crystalline two- or threedimensional
structure~\cite{drewsen1998large,Birkl}. Especially interesting are
planar ion crystals, where usually one of the confining
radial potentials is much tighter than the axial one, as
such structures with nondegenerate radial frequencies do
not rotate~\cite{Hasegawa,Taylor} and allow to address and observe individual
ions. Recent proposals have outlined how to achieve
laser-induced spin-spin interactions on such 2D ion crystals
and how to use their spatial arrangement for the
realization of spin lattices exhibiting frustration~\cite{BERMUDEZ,bermudez2012}.
It was also proposed how to study the spin-phonon interactions
coupling the geometric structure of the ion chain to
the spin-spin interaction of the chain to realize a Peierls
instability~\cite{Plenio} or the Jahn-Teller quantum phase transition~\cite{PORRAS1,PORRAS}. Topological defects in the zigzag configuration
were proposed for simulation of quantum effects with
solitons~\cite{soliton}. It was also proposed that the dynamics of
the structural transition from linear to zigzag configuration
may be induced by electronic excitations or a fast change
of the trap parameters~\cite{ulm,giovanna,ulmsaar,ulmsaar1} and it would allow for the
verification of the predicted scaling laws for defect formation
when traversing the transition, that is the Kibble-Zurek
mechanism~\cite{Kibble,Zurek}. Furthermore, the double well structure
realized by the two different configurations of the zigzag
ion crystals can be manipulated and may, thus, allow for
the creation of a coherent superposition between these two
configurations and hence, serve as a test bed for decoherence
models~\cite{ulm,PhysRevLett.108.023003}.%~\cite{leibfried2003,britton2012,islam}.

Spectroscopy and quantum simulation experiments with
planar ion crystals in Penning traps have recently been
reported~\cite{britton2012,sawyer2012spectroscopy}. However, no experiments have been successful
in using planar crystals in a Paul trap. This is due to
the high complexity of controlling such crystalline twodimensional
structures. Required is the knowledge of
eigenmodes and eigenfrequencies of such crystals, since
it is important for the cooling and for the design of ion-ion
interactions which rely on the setting of laser parameters.
These are tailored to induce spin-dependent light forces via
Stark effects~\cite{leibfried2003,britton2012,islam}.

Here, we present a combined experimental and theoretical
study of eigenmodes of ions in planar crystals. We
describe an accurate experimental determination of positions
of ions and the frequencies of the eigenmodes of the
crystal structures together with a comparison of these data
with theoretical expectations based on pseudopotential
approximation (PPT) and a full dynamical classical theory
for solving the linearized Coulomb problem in ion crystals
using a Floquet-Lyapunov approach (FLT)~\cite{landa2012ions,landa2012modes}. From
observations of multiple sequential phase transitions
between different structures of the ion crystal, we map out
the phase diagram and find good agreement with theoretical
predictions following PPT for both the transition points
and for the positions of individual ions in the different
phases. The measurement of the eigenfrequencies relies
on sideband spectroscopy and is applied here to the vibrational
frequencies of a three-ion zigzag crystal. Surprisingly,
some modes exhibit a significant 37 kHz deviation when
compared to calculations in PPT. We can understand the
measured data quantitatively only if the full time-dependent
solution of the trapping potential is taken into account, with
the eigenfrequencies calculated using FLT.

The positions of the ions in a crystal are determined by
the mutual Coulomb repulsion together with a static electric
potential in the z direction and a dynamic radial
trapping potential. A single charged particle in the timedependent
potential of a quadrupole Paul trap, or the
center-of-mass (c.m.) modes of a general crystal of ions,
obey decoupled, linear Mathieu equations of motion in
each spatial direction~\cite{leibfried2003quantum}. The rf and dc trap voltages,
together with the rf-frequency $\Omega$ and the mass of the ions
m, determine the dimension-free Mathieu parameters in
each direction of space $i\in\left\{x,y,z\right\}$ as $a_i=4 e U_{DC}/ \gamma_i m \Omega^2$
and $q_i=2 e U_{\text{RF}}/ \gamma_i' m \Omega^2$, where $\gamma_i$ and $\gamma_i'$, where $\gamma_i$ and $\gamma_i'$
i are geometrical
factors denoting the curvature of the respective potentials
and $\xi=\Omega t/2$.
The Mathieu equation is solved
\begin{equation}
    \frac{d^2y}{d\xi^2} + [a-2q~ cos(2\xi)]y = 0. \label{Eq:math}
\end{equation}
One derives stable solutions with characteristic exponents
 $\beta_i\approx\sqrt{a_i + q^2_i/2}$, and obtains a harmonic time-independent
pseudopotential with frequencies $\omega_i=\beta_i\Omega/2$.

In the pseudopotential approximation, the ions arrange
in positions corresponding to a stable minimum of the
time-independent potential, the determination of which
follows the method for linear crystals~\cite{james1998quantum}, extending it
to three dimensions. In the time-dependent potential, the
equivalent of a minimum configuration crystal is a periodic
solution with the ions oscillating at the rf-frequency $\Omega$,
about well-defined average locations (the oscillation
denoted as micromotion). The ion coordinates in the radial
direction obey
\begin{equation}
    y_n(t) = \bar{y}_n[1-\frac{q_y}{2}\cos(\Omega t)] + {\rm O}\left(\frac{q_y^2}{4}\right), \label{Eq:y}
\end{equation}
while in the axial z direction, the micromotion is negligible~\cite{landa2012ions}.

In the experiment, the resonance fluorescence of
Doppler cooled $^{40}$Ca$^+$ ion crystals near 397 nm is imaged
on a CCD camera~\cite{nagerl1998ion}. We use a linear Paul trap consisting
of four cylindrical rods (2$\times$ rf and 2$\times$ dc, of diameter
2.5 mm and at diagonal distance between centers of
4.7 mm), and two end caps at 10 mm distance. Note that
the degeneracy of the radial modes is lifted using a dc
offset voltage of about 0.5 V. We operate with a rf amplitude
off $U_{\text{RF}} \sim 300~\text{V}_{\text{pp}}$ at $\Omega/(2\pi) = 14.62$~MHz MHz
and end cap voltages of 350 V, yielding secular frequencies
$\omega_{y,z}/(2\pi)$=(316, 111)~kHz and a much larger $\omega_{x}$. All
secular frequencies are determined from the resonant electric
excitation of the c.m. modes~\cite{naegerl1998coherent}.

\begin{figure}[ht]
\includegraphics[width=0.95\linewidth]{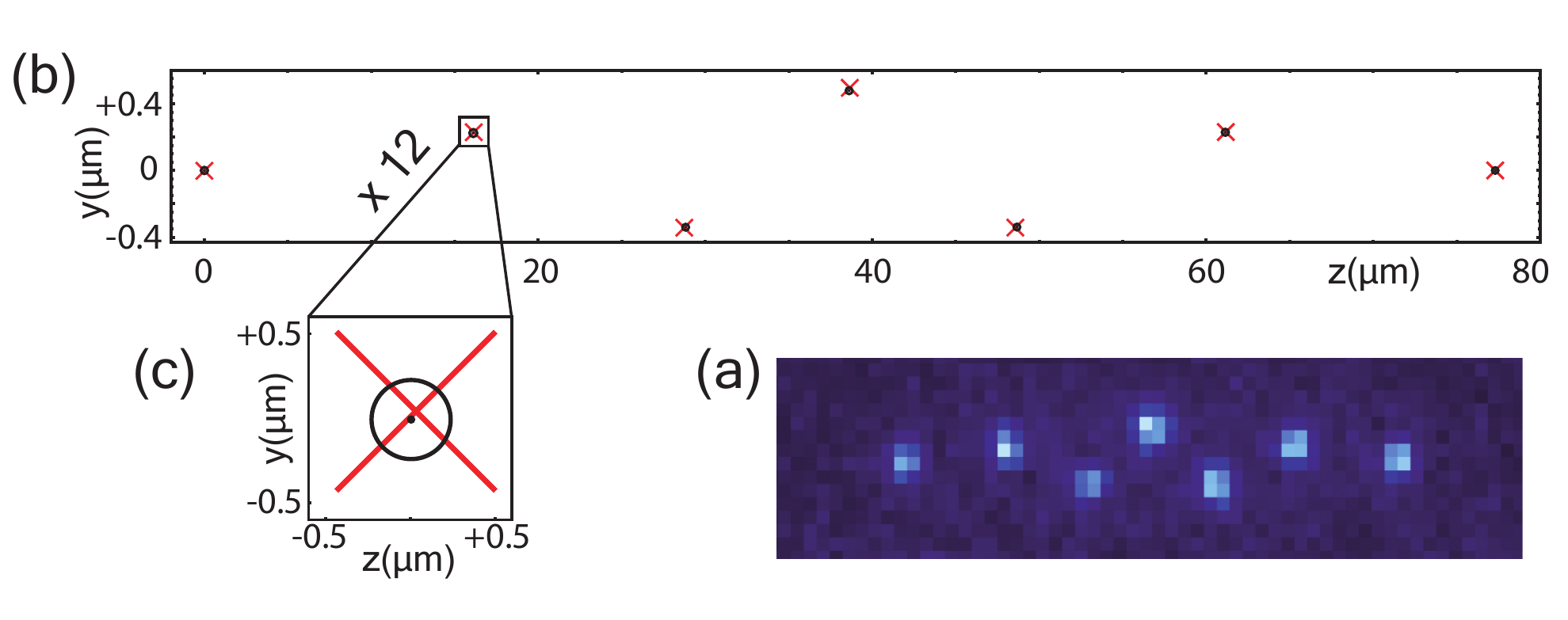}
\caption{(color online). Determination of ion positions in planar
crystals: (a) The ion fluorescence near 397 nm is imaged on a
CCD chip. (b) The ion positions (black dots) are determined by
averaging over 100 exposures and compared with the result of a
numerical simulation assuming Coulomb repulsion in a harmonic
trap pseudopotential (red crosses). The experimental
data allow for a precision of 50 nm as indicated by the circle
for a 1 $\sigma$ standard deviation.}
\label{Fig:positions}
\end{figure}

\emph{Equilibrium positions}.$-$Are determined from CCD images
such as shown in Fig. \ref{Fig:positions} with a 7-ion crystal.
Averaging over 100 exposures and applying a Gaussian
fit to the data, we determine the ion locations, see Fig. \ref{Fig:positions}(b).
As the planar crystal is observed at an angle of 45$^{\circ}$, the y
axis is compressed by $\sqrt{2}$. The magnification of the imaging
system is determined from the c.m. mode frequency in
z direction and the z distance in a linear two-ion crystal.
The observed fluorescence precisely indicates the ion
positions. The measured data show perfect agreement with
the theoretical expectations following PPT at the level of a
few parts in 10$^5$, see Fig. \ref{Fig:positions}(c), for a crystal of 7 ions, and
similar agreement is found for crystals with 6 to 17 ions.

\begin{figure}[ht]
\includegraphics[width=0.95\linewidth]{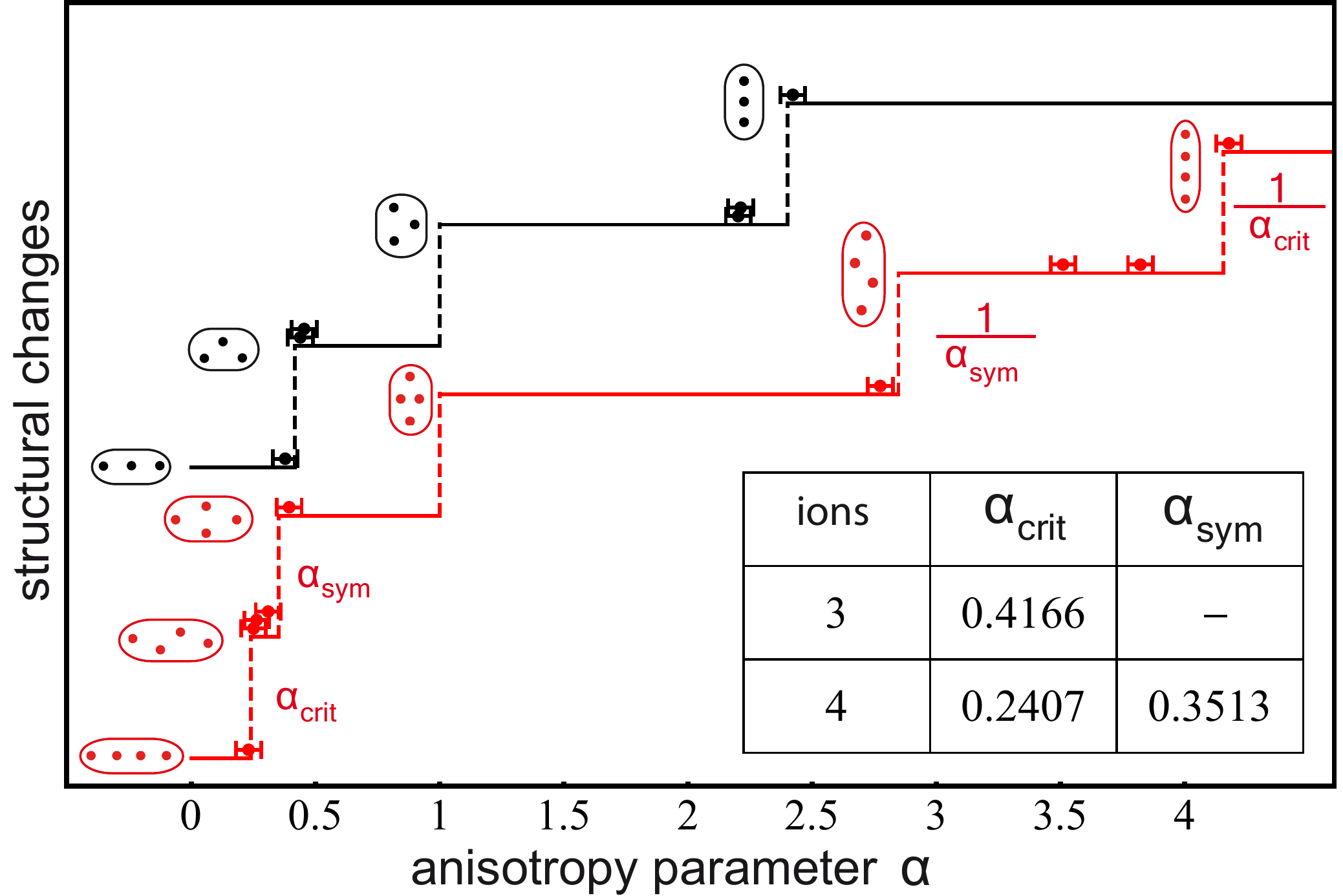}
\caption{(color online). Phase transitions for a three-ion (black)
and a four-ion (red) crystal. Theoretical (PPT) critical $\alpha's$ are
indicated by vertical dashed lines with the corresponding crystal
configurations. Experimental data are plotted at those $\alpha$ where a
certain configuration is observed, with an error of 0.05 in $\alpha$. For
an ion crystal with an even number of ions (here, N=4), $\alpha_{sym}$ is
the relevant parameter, where the structure symmetry changes.
The measurements are taken in a linear micro trap with
$\Omega/(2\pi)$~=~22.7~MHz, U$_{\text{RF}} \sim$ 300~$\text{V}_{\text{pp}}$ yielding frequencies
$\omega_{\{y,z\}}/(2\pi)$~=~(0.421, 0.626)~MHz and a much larger $\omega_{x}$. Inset:
Table of calculated critical $\alpha$.}
\label{Fig:transitions}
\end{figure}

\emph{Structural phase transitions}.$-$Are induced when the
value of $\alpha$ is varied. While so far the first linear-to-zigzag
transition was investigated~\cite{losalamos}, we observe multiple consecutive
critical $\alpha$, see Fig. \ref{Fig:transitions}. The experimentally determined
positions and the critical values of $\alpha$ agree with the
prediction of PPT. Comparing the PPT prediction and the
solution of the full time-dependent equations, with $q\lesssim 0.1$, the relative differences for critical $\alpha$ are only about
$1\%$, not resolved in the experimental data.

\emph{Normal mode frequencies}.$-$Result from the Coulomb
forces between ions at their equilibrium positions. For the
calculation in PPT, these forces are expanded in small
excursions about the equilibrium positions, linearized, and
the Hessian matrix is solved. Laser spectroscopy provides a
precise tool to determine eigenfrequencies with a relative
accuracy of 0.2 percent or better, and we do not find agreement
between the measurements and the PPT calculation.
The corresponding values for mode frequencies for the
three-ion crystal under study, in a zigzag configuration with
$\alpha \approx$ 0.53, are in Table \ref{Table1}. Only recently, the influence of
micromotion on the frequencies of secular modes has been
investigated theoretically, which results in significant corrections
as compared to PPT, even for relatively small $q$ values.

\begin{figure}[ht]
\includegraphics[width=0.95\linewidth]{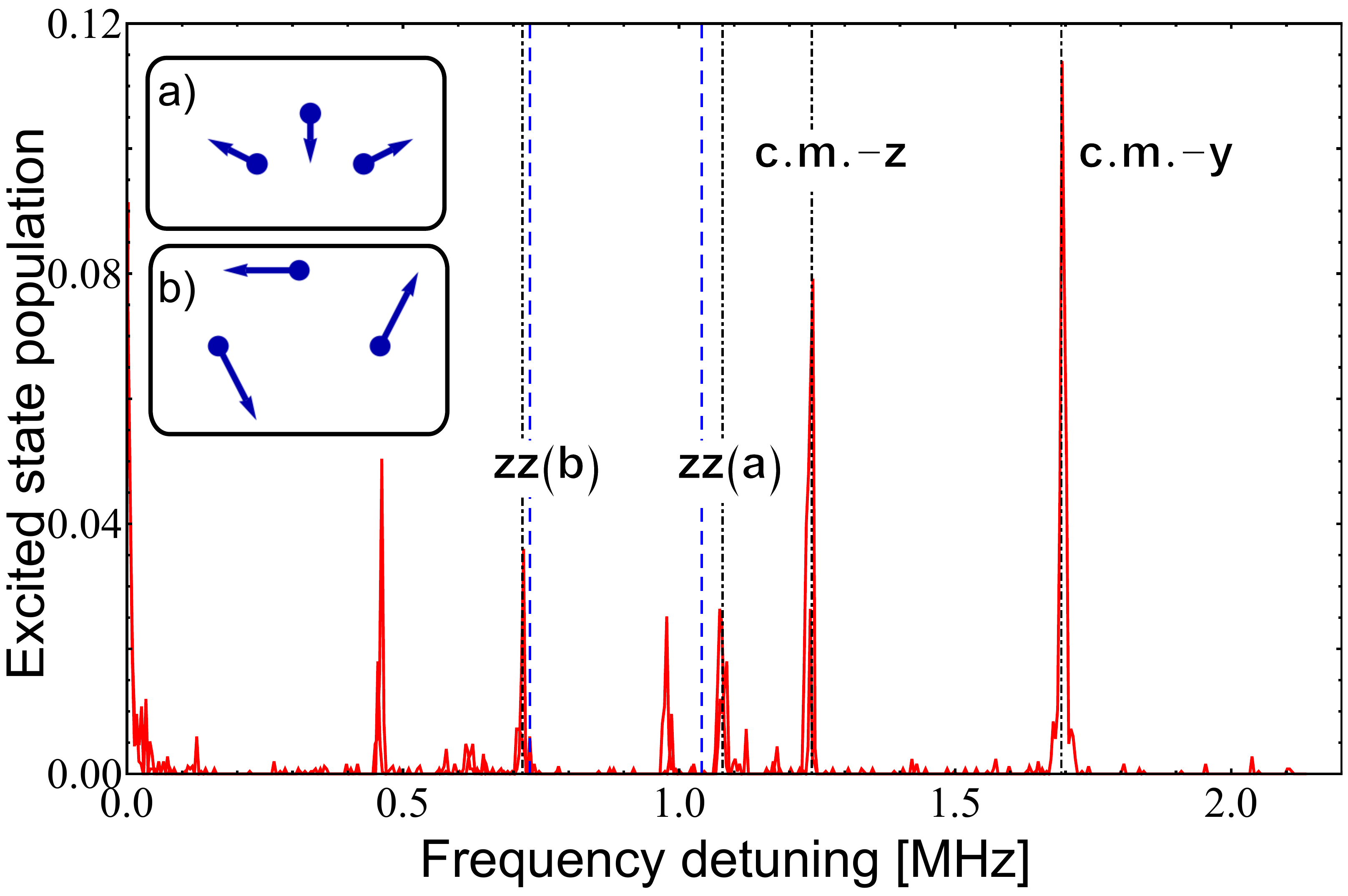}
\caption{(color online). Spectrum of vibrational modes in a three-ion
crystal. The excitation from the S$_{1/2}$ to D$_{5/2}$ state is plotted as
the laser frequency near 729 nm is scanned over the resonance
between a frequency detuning of $\pm$ 2~MHz. We observe excitation
at the frequencies of red and blue motional sidebands, symmetrical
around the carrier. The spectrum is plotted versus the modulus
of frequency detuning, such that red and blue sidebands fall
on each other. Measurements are taken in a linear micro trap
with $\Omega/(2\pi)$~=~35.07~MHz and U$_{\text{RF}} \sim$ 420~$\text{V}_{\text{pp}}$ yielding frequencies
$\omega_{\{x,y,z\}}/(2\pi)$~=~\{2940(10), 1695(3), 1238(2)\}~kHz. We
identify zigzag modes near (a) 1078(2) kHz and (b) 0.714
(2) kHz. The blue dashed lines indicate the expected frequencies
for zigzag modes from PPT, while the black dashed-dotted
lines show the outcome of the calculation of FLT. Insets display
eigenvectors of zigzag modes (a) and (b). The other resonances
correspond to mixing frequencies of two normal mode
frequencies.}
\label{Fig:Spectrum}
\end{figure}

Table \ref{Table1} displays the resulting frequencies of the zz
modes. Experimental uncertainties result from fluctuations
of the trap control voltages, drifts of the laser reference
cavity, or magnetic field noise during the scan. We obtain
the errors from a statistics of a comparison of red and blue
sideband frequencies in many scans like that plotted in
Fig. \ref{Fig:Spectrum}, obtained on the same day and under identical
conditions as the data. The PPT prediction for the zz
frequencies is calculated based on the experimental data
for the $\omega_{z,y}$ which are c.m. modes (whose frequencies are
the same in PPT approximation and full dynamic theory).
Systematic errors, including the ac Stark effect $<$1 kHz,
are contained in the error budget. From the z and y mode
frequencies, we first determine the ion positions in the
pseudopotential and then the zz mode frequencies.
Experimental uncertainties in $\omega_z$ and $\omega_y$ lead to uncertainties
for the PPT zigzag mode frequency prediction of 14
and 12 kHz, respectively. Theory values for FLT are
obtained as a best fit to all experimentally determined
frequencies~\cite{supplemental}.

The breakdown of PPT results from the fast oscillations
of ions about their equilibrium positions at the rffrequency,
as in Eq. (\ref{Eq:y}), which modifies periodically the
Coulomb forces between ions. Thus, it is not justified to
assume static positions for the Hessian matrix. The forces
between the ions can be expanded in a Fourier series, and
the resulting equations can be solved in terms of decoupled
modes~\cite{landa2012ions}, which describe oscillations with secular frequencies,
superimposed on micromotion at the rf-frequency.
This effect resembles the Lamb shift where
the Zitterbewegung of the bound electron leads to a modification
in the hydrogen energy levels~\cite{WEITZ}. A short
description of the method of solution, together with code
for the calculations presented in this Letter, is available in
the Supplemental Material~\cite{supplemental}.

\begin{table}{}
\centering
\caption{Frequencies of a three-ion zigzag crystal in units of
kHz$/(2\pi)$}
\begin{tabularx}{\columnwidth}{X X X X X}
\hline\hline
& $\omega_{zz(b)}$ & $\omega_{zz(a)}$ & $\omega_{z} $ & $\omega_{y}$ \\
\hline
Exp. & 714(2) & 1078(2) & 1238(2) & 1695(3) \\
PPT & 730(14) & 1041(12) &  &  \\
FLT & 715.1 & 1078.5 & 1239.5 & 1690.7\\ \hline\hline
\end{tabularx}
 \label{Table1}
\end{table}

We test the FLT prediction experimentally using a threeion
crystal and performing resolved sideband spectroscopy
on the narrow $S_{1/2}$ to $D_{5/2}$ transition near 729~nm. In the
spectrum (Fig. \ref{Fig:Spectrum}), we identify the vibrational frequencies
of a three-ion zigzag crystal. When testing PPT, we use the
c.m. modes in axial and radial y direction to generate
predictions for the zigzag frequencies, which deviate
from the experimental values by 37 and 15 kHz, respectively.
For FLT, we fit the five measurements of the three
c.m. modes and the two lowest planar zigzag modes using a
weighted-least-squares fit, with three parameters $q_y$, $a_z$ and $a_y$ (imposing $a_x=-a_z-a_y$ as required from the
Laplace equation). The weighted-least-squares fit norm is
a random variable distributed like $\chi^2$ with two degrees of
freedom, see Ref.~\cite{supplemental}. The theoretical values coincide
exactly with both the eigenfrequencies, fitting the data
with about $22\%$ probability with all the eigenfrequencies.
A similar fitting procedure for the PPT results in a negligible
probability of the order $10^{-11}$, reflecting that PPT
frequencies do not agree with the experimental finding.

\begin{figure}[ht]
\includegraphics[width=0.95\linewidth]{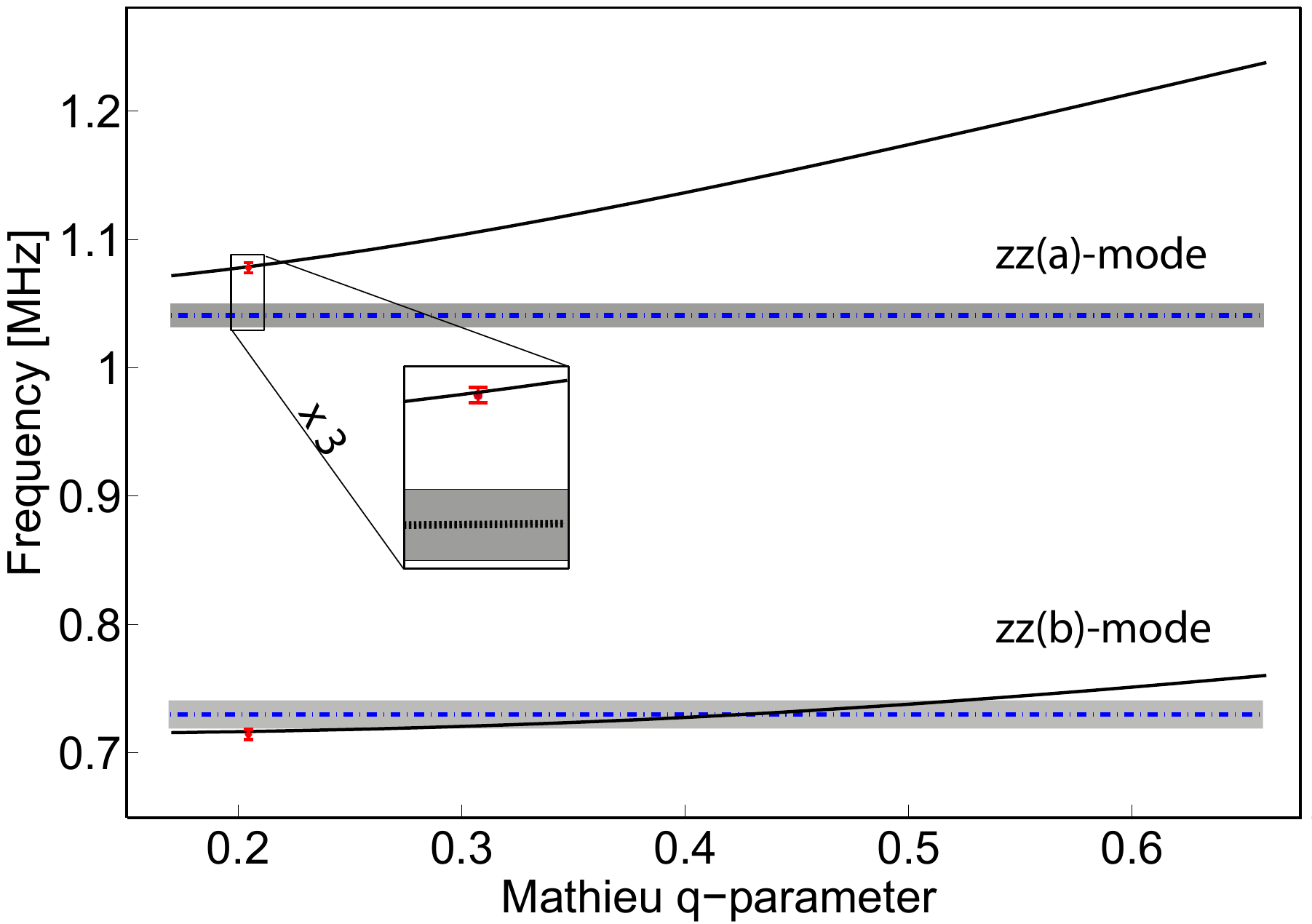}
\caption{(color online) The calculated eigenfrequencies for the zigzag modes (a) and (b) in a three-ion crystal plotted versus the Mathieu parameter $q$, with $a$-parameters adjusted such that the c.m. frequencies are constant. The $zz$ eigenfrequencies show a slope with $q$. The dashed line results from the PPT, see Table \ref{Table1}, including the prediction error (grey shaded). Experimental data point fit the theoretical FLT expectation near $q \approx 0.2$, where $\Omega/(2\pi)$ equals 35.07~MHz. Inset: with zoomed part showing the fit of the FLT and the 3$\sigma$ deviation of data against PPT.}
\label{Fig:zigzagShift}
\end{figure}

The strong influence of the trap drive on the eigenfrequencies
of zigzag modes is further explained by a theoretical
analysis, where the Mathieu parameter $q$ is varied
and the variation of the mode frequencies $zz(a)$ and $zz(b)$ is
plotted, see Fig. \ref{Fig:zigzagShift}. In the experiment, the $q$ value is fixed
and can not varied easily over a large range. The data
points, near $q=0.2$, for the $zz(a)$ and $zz(b)$ mode, hit
the FLT prediction and exclude the PPT frequencies. The
fractional frequency shift reaches, for high $q$ values, a level
of up to 20$\%$. Examining Fig. \ref{Fig:Spectrum}(a), it is understandable
why the $zz(a)$ mode is affected the most by the micromotion,
and why its frequency increases monotonically
with $q$. It can be seen from Eq. (\ref{Eq:y}) that the micromotion
amplitude of each ion is proportional to the negative of its $y$
equilibrium position, so that this motion has a large projection
on the eigenvector of the $zz(a)$ mode. At each point along the
periodic trajectory, the restoring forces are larger for this
mode than at the center, and thus, its frequency increases
with the amplitude of the micromotion, hence, with $q$.

{\em Conclusion. ---}We calculated and characterized experimentally
the behavior of two-dimensional crystals under
the influence of micromotion. We show that while the
pseudopotential theory fails to explain the experimental
results to an accuracy within 3$\sigma$-deviation, the newly
established method predicts the experimental finding correctly.
We could show that FLT provides an experimentally
verified tool for understanding the static and dynamic
properties of mesoscopic Coulomb crystals and might
prove to be a key ingredient for establishing these systems
as a new experimental platform. The presented results are
an essential prerequisite for the success of quantum simulation
in two-dimensions where the effective Hamiltonian
depends on the crystal frequencies of all normal modes in
the presence of micromotion.

It is important to note that micromotion is a driven
motion with a well-defined phase: Therefore, it does not
delocalize the ion wave packet, leading to a larger effective
Lamb Dicke parameter. The micromotion merely acts as to
affect the strength of atom-laser interactions via phase
modulation. It does not prevent from exploiting spin dependent
forces for quantum simulation schemes.

Based on our results, a theory could be constructed that
takes into account such effects in the effective Hamiltonian.
Moreover, the precise knowledge of the behavior of the
crystal under the influence of micromotion is crucial for
quantum-to-classical simulation experiments that use such
transitions in condensed matter or high energy effects, such
as the Kibble Zurek mechanism. The deviation of critical $\alpha$
values from the pseudopotential predictions is expected to
be larger for larger values of $q$ and merits further experimental
and theoretical investigation.

We acknowledge financial support by the European commission within the Integrated Projects AQUTE and QESSENCE, the EU STREP PICC, the German-Israeli-Foundation and the Alexander von Humboldt Foundation.

\bibliographystyle{apsrev4-1}
\bibliography{zigzagModes}

\end{document}